\documentclass[aps,preprint,superscriptaddress,nofootinbib,amsmath,amssymb]{revtex4}
\usepackage{bm}
\usepackage{graphicx,color}
\usepackage{subfigure}
\def\be{\begin{eqnarray}}
\def\ed{\end{eqnarray}}
\def\non{\nonumber}

\begin{document}

{\begin{flushright}{IPMU15-0060}
\end{flushright}}

\title{Dark matter for excess of AMS-02 positrons and antiprotons}

\author{Chuan-Hung Chen}
\email[e-mail: ]{physchen@mail.ncku.edu.tw}
\affiliation{Department of Physics, National Cheng-Kung
University, Tainan, Taiwan 70101, R.O.C.}

\author{Cheng-Wei Chiang}
\email[e-mail: ]{chengwei@ncu.edu.tw}
\affiliation{Center for Mathematics and Theoretical Physics and Department of Physics, 
National Central University, Taoyuan, Taiwan 32001, R.O.C.}
\affiliation{Institute of Physics, Academia Sinica, Taipei, Taiwan 11529, R.O.C.}
\affiliation{Physics Division, National Center for Theoretical Sciences, Hsinchu, Taiwan 30013, R.O.C.}
\affiliation{Kavli IPMU (WPI), UTIAS, The University of Tokyo, Kashiwa, Chiba 277-8583, Japan}

\author{Takaaki Nomura}
\email[e-mail: ]{nomura@mail.ncku.edu.tw}
\affiliation{Department of Physics, National Cheng-Kung
University, Tainan, Taiwan 70101, R.O.C.}


\date{\today}

\begin{abstract}
We propose a dark matter explanation to simultaneously account for the excess of antiproton-to-proton and positron power spectra observed in the AMS-02 experiment while having the right dark matter relic abundance and satisfying the current direct search bounds.  We extend the Higgs triplet model with a hidden gauge symmetry of $SU(2)_X$ that is broken to $Z_3$ by a quadruplet scalar field, rendering the associated gauge bosons stable weakly-interacting massive particle dark matter candidates.  By coupling the complex Higgs triplet and the $SU(2)_X$ quadruplet, the dark matter candidates can annihilate into triplet Higgs bosons each of which in turn decays into lepton or gauge boson final states.  Such a mechanism gives rise to correct excess of positrons and antiprotons with an appropriate choice of the triplet vacuum expectation value.  Besides, the model provides a link between neutrino mass and dark matter phenomenology.
\end{abstract}

\maketitle

\section{Introduction}

After the discovery of 125-GeV Higgs boson with many properties consistent with the standard model (SM) expectations~~\cite{:2012gk,:2012gu}, we are left with two pieces of empirical evidence that call for new physics explanations.  One is the phenomenon of neutrino oscillations that leads to the question about how neutrino mass is generated.  The other is astronomical observation of gravitational effects caused by dark matter (DM), whose abundance is bound to be about 5 times that of ordinary matter.

In the SM, masses of quarks and massive gauge bosons are generated through Yukawa and gauge couplings with the condensate of the Higgs doublet field through the Brout-Englert-Higgs (BEH) mechanism~\cite{Englert:1964et, Higgs:1964pj}.  One minimal extension for giving tiny mass to neutrinos is done by introducing a complex Higgs triplet through the type-II seesaw mechanism, the so-called Higgs triplet model (HTM)~\cite{Magg:1980ut,Lazarides:1980nt,Mohapatra:1980yp,Ma:1998dx,Konetschny:1977bn,Schechter:1980gr,Cheng:1980qt,Bilenky:1980cx}.  In this case, the vacuum expectation value (VEV) of the Higgs triplet field is induced by electroweak symmetry breaking (EWSB) and controls whether the charged Higgs bosons derived from the triplet field decay dominantly to leptons or weak gauge bosons.  Direct searches of the doubly charged Higgs boson predicted in this model at the Large Hadron Collider (LHC) generally put a lower mass bound at about 400 GeV~\cite{Akeroyd:2010ip,Chiang:2012dk,Chatrchyan:2012ya,ATLAS:2012hi}.

In the past few years, many experiments have reported indirect evidence of DM, such as the excess of positron fraction observed by PAMELA~\cite{Adriani:2008zr},  Fermi-LAT~\cite{FermiLAT:2011ab} and AMS-02~\cite{Aguilar:2013qda}, the excess of positron+electron flux observed by ATIC~\cite{Chang:2008aa}, HESS~\cite{Aharonian:2008aa, Aharonian:2009ah}, Fermi-LAT~\cite{Ackermann:2010ij}, PAMELA~\cite{Adriani:2011xv} and AMS-02~\cite{Aguilar:2013qda}, the excess of gamma-ray spectrum at the Galactic Center~\cite{Daylan:2014rsa,Zhou:2014lva,Calore:2014xka,Abazajian:2014hsa,Calore:2014nla}, and so on.  The AMS Collaboration also confirms with an unprecedented precision the excess of positron fraction in the energy range of $[0.5, 500]$ GeV~\cite{Accardo:2014lma}, the positron+electron flux from 0.5 GeV to 1 TeV~\cite{Aguilar:2014fea}, and a deviation of the antiproton fraction from secondary astrophysical sources of cosmic ray collisions for the antiproton kinetic energy between $50 - 500$ GeV~\cite{AMS2:2015}.  Although still uncertain whether the observed antiproton spectrum is still consistent with the background of secondary antiprotons~\cite{Giesen:2015ufa,Jin:2015sqa}, several studies~\cite{Ibe:2015tma,Hamaguchi:2015wga,Lin:2015taa} have attempted to explain the possible excess of antiproton using DM annihilations and/or decays.  It is well-known that the excess of both electron and position fluxes require additional contributions to the thermally averaged DM annihilation cross section~\cite{Chen:2014lla,Belotsky:2014haa} than is required by the antiproton fraction spectrum.  
In this work, we propose a weakly-interacting massive particle (WIMP) DM model that readily accommodates the two sets of data.

In this work, we extend the HTM by having an additional hidden $SU(2)_X$ symmetry that is broken by a scalar quadruplet down to a residual $Z_3$ symmetry {\it a la} Krauss-Wilczek mechanism~\cite{Krauss:1988zc, Hambye:2008bq,Chiang:2013kqa,Khoze:2014woa,Baek:2013dwa,Chen:2015nea}, making the associated gauge bosons our DM candidates.  The quadruplet also has interactions with both the SM doublet and the Higgs triplet and plays the role of mediator that connects the hidden sector and the visible sector of SM particles and exotic Higgs bosons.  A model also invoking a scalar triplet field but using a singlet scalar as a cold DM candidate was proposed in Ref.~\cite{Dev:2013hka}.  As a result, the DM can annihilate into a pair of exotic Higgs bosons that in turn decay into leptons or weak gauge bosons as alluded to before.  The leptonic channel will lead to production of electrons and positrons, and the gauge channel to protons and antiprotons after hadronization.  With an appropriate choice of the triplet VEV, the model can explain simultaneously the observed spectra of positron and antiproton fluxes.  It should be emphasized that the choice of $SU(2)_X$ gauge group is just to provide an explicit model with stable DM candidates.  Other DM models from a hidden sector that couples mainly to the triplet Higgs fields should catch the same features as well.

\section{The Model}

In this model, we extend the SM gauge group by an additional $SU(2)_X$ symmetry with the associated gauge field denoted by $X_\mu^a$.  This symmetry is broken by an $SU(2)_X$ quadruplet field $\Phi_4= (\phi_{3/2}, \phi_{1/2}, -\phi_{-1/2}, \phi_{-3/2})^T/\sqrt{2}$ that does not carry SM gauge charges, where the subscript stands for the eigenvalue of $T_3$, the third $SU(2)_X$ generator, of the field and $\phi_{-i} = \phi^*_{i}$.  Finally, we also introduce a complex Higgs field $\Delta$ that is a triplet under the SM $SU(2)_L$ and carries hypercharge $Y = 1$.  Here we adopt the convention that the electric charge $Q = T_3 + Y$.

More explicitly, the complete Lagrangian invariant under the $SU(2)_X \times SU(2)_L \times U(1)_Y$ gauge group is
\begin{align}
{\cal L}=& 
{\cal L}_{SM} 
+ (D_\mu \Delta )^\dagger D^\mu \Delta- \left[  L^T C {\bf y^\ell }  i\sigma_2 \Delta P_L L +h.c. \right] 
\non \\
&+ \left( D_\mu \Phi_4\right)^\dagger D^\mu \Phi_4 - V(\Phi,\Delta, \Phi_4) 
- \frac{1}{4} X^a_{\mu\nu} X^{a \mu \nu} ~,
\label{eq:lang}
\end{align}
where ${\cal L}_{SM}$ is the Lagrangian of SM, $L$ denotes the left-handed lepton doublet field, $C$ is the charge conjugation, ${\bf y^\ell}$ is the Yukawa coupling matrix, $\Phi$ is a Higgs doublet, and the field strength tensor $X^a_{\mu \nu} = \partial_\mu X^a_\nu -\partial_\nu X^a_\mu - g_X (\vec X_\mu \times \vec X_\nu)^a$.  The most general Higgs potential
\begin{align}
V(\Phi,\Delta, \Phi_4) =&
\mu^2 \Phi^\dagger \Phi + \lambda (\Phi^\dagger \Phi)^2 + m^2_\Delta Tr\Delta^\dagger \Delta 
+ \lambda_\Delta (Tr \Delta^\dagger \Delta)^2 + \bar \lambda_\Delta (Tr\Delta^\dagger \Delta)^2  
\non \\
&+ \mu^2_\Phi \Phi^\dagger_4 \Phi_4 + \lambda_\Phi (\Phi^\dagger_4 \Phi_4)^2 
+ \mu_\Delta \left(\Phi^\dagger i\tau_2 \Delta^\dagger \Phi + {\rm h.c.} \right) 
+ \lambda_1 \Phi^\dagger \Phi Tr\Delta^\dagger \Delta
\non \\
&+ \lambda_2 \Phi^\dagger \Delta \Delta^\dagger \Phi 
+ \bar\lambda_2 \Phi^\dagger \Delta^\dagger \Delta \Phi 
+ \lambda_3  \Phi^\dagger_4 \Phi_4 \Phi^\dagger \Phi 
+ \lambda_4  \Phi^\dagger_4 \Phi_4 Tr\Delta^\dagger \Delta ~, 
\label{eq:vphi4}
\end{align}
where, as in the SM, $\mu^2 < 0$ is required for the EWSB, $\Phi = (G^+, (v+\phi + i G^0)/\sqrt{2})^T$ is the SM Higgs doublet field, and the scalar triplet of $SU(2)_L$ is written in an $SU(2)_L \times SU(2)_R$ covariant form as
\begin{align}
\Delta = \begin{pmatrix}
    \delta^+/\sqrt{2} & \delta^{++}  \\ 
    (v_\Delta + \delta^0 + i\eta^0)/\sqrt{2} & -\delta^+/\sqrt{2} \\   
\end{pmatrix} ~,
\end{align}
where the triplet VEV $v_\Delta$ is induced by the doublet VEV through the $\mu_\Delta$ term and is constrained by the electroweak rho parameter to be less than about 1 GeV.  In fact, to produce the right positron and antiproton spectra given by the AMS-02 experiment, $v_\Delta$ is required to be ${\cal O}$(10-100) keV.  With $m_\Delta$ assumed to be of ${\cal O}$(TeV), $\mu_\Delta$ is also about the same scale as $v_\Delta$.

To break the $SU(2)_X$ symmetry while preserving a discrete symmetry to stabilise DM candidates, we require $\mu_\Phi^2 < 0$ so that the $\Phi_4$ field spontaneously develops a VEV in the $T_3 = \pm 3/2$ component:~\cite{Chen:2015nea}
\begin{align}
\phi_{\pm 3/2} = \frac{1}{\sqrt{2}} \left( v_4 + \phi_r \pm i \xi \right) ~,
\label{eq:vev}
\end{align}
where $v_4$ is assumed to be ${\cal O}(10)$ TeV.  As a result, the $SU(2)_X$ gauge bosons $\chi_\mu ~(\bar \chi_\mu)= (X^1_\mu \mp i X^2_\mu)/\sqrt{2}$ and $X^3_\mu$ acquire their masses, $m_{\chi} =\sqrt{3} g_X v_4 /2$ and $m_{X^3}=\sqrt{3} m_\chi$, at the TeV scale after absorbing the $\phi_{\pm 1/2}$ and $\xi$ as their longitudinal components.  Moreover, there is a residual $Z_3$ symmetry in the model, under which $\chi_\mu$ and $\bar \chi_\mu$ carry nonzero charges and serve as good candidates of DM.  Finally, the physical states in the hidden sector and the quadruplet are $\chi_\mu$, $\bar \chi_\mu$, $X^3_\mu$ and $\phi_r$, where $\phi_r$ plays the role of messenger between the hidden sector and the visible sector.

We now work out the relevant couplings for our analysis.  From the kinetic term of $\Phi_4$ and the breaking pattern of Eq.~(\ref{eq:vev}), the gauge interaction between $\phi_r$ and $\chi_\mu$ is given by~\cite{Chen:2015nea}
\begin{align}
I_{\chi \bar\chi \phi_r} =& \sqrt{3} g_X m_\chi \phi_r \chi_\mu \bar\chi^\mu ~,
\label{eq:igg}
\end{align}
where, as mentioned above, $m_\chi \sim {\cal O}$(TeV).  In general, the neutral components in $H$, $\Delta$, and $\Phi_4$ can mix under the mass matrix
\begin{align}
M^2 = 
\begin{pmatrix} 
M_\phi^2 & \lambda_3 v v_4 & -\sqrt{2} \mu_\Delta v 
\\ 
\lambda_3 v v_4 & M_{\phi_r}^2 & \lambda_4 v_\Delta v_4 
\\ 
-\sqrt{2} \mu_\Delta v & \lambda_4 v_\Delta v_4 & M_\Delta^2 
\end{pmatrix}
\label{eq:mass}
\end{align}
in the basis of $(\phi,\phi_r,\delta^0)$, where $M_\phi = \sqrt{2 \lambda} v$, $M_{\phi_r} = \sqrt{2 \lambda_\Phi} v_4$, and $M_\Delta^2 = m_\Delta^2+(\lambda_1 + \lambda_2+ \bar \lambda_2)v^2/2 + \lambda_4 v_4^2/2$.  As mentioned above, $\mu_\Delta$ and $v_\Delta$ are much smaller than the other mass parameters in Eq.~(\ref{eq:mass}).  Therefore, only $\phi$ and $\phi_r$ mix to render the mass eigenstates
\begin{align}
\begin{pmatrix}
    h \\ 
    H \\ 
\end{pmatrix} =
\begin{pmatrix}
    \cos\theta & \sin\theta \\ 
    -\sin\theta & \cos\theta \\ 
\end{pmatrix}
\begin{pmatrix}
    \phi \\ 
    \phi_r \\ 
\end{pmatrix} ~,
\label{eq:mixing}
\end{align} 
where $h$ denotes the SM-like Higgs boson of mass 125 GeV, $H$ is the other physical Higgs boson with a mass at the TeV scale, and the mixing angle $\theta$ given by $\tan2\theta = 2\lambda_3 v v_4 /(M^2_{\phi_r} - M^2_\phi)$ is taken to be small in view of the hierarchy between $v$ and $v_4$ and by assuming small $\lambda_3$.  In addition, we require $\lambda_{1,2}$ and $\bar\lambda_2$ to be sufficiently small so that the mass of $\delta^0$ is smaller than $m_\chi$.  Finally, we obtain the following approximate formulas for the masses of physical scalar bosons:
\begin{align}
&
m_h \approx  m_\phi = \sqrt{2 \lambda}\, v ~,~~~ 
m_{H} \approx m_{\phi_r}= \sqrt{2 \lambda_\Phi}\, v_4 ~,
\non \\
&
m_{\delta^{\pm\pm} } \approx m_{\delta^{\pm}} \approx m_{\delta^0}= m_{\eta^0} 
= M_\Delta ~.
\end{align}
In the limit of vanishing $\lambda_3$, $H = \phi_r$ does not couple with SM particles directly.  Now the only important parameter that controls the phenomenology of DM interactions with the visible sector is $\lambda_4$.  Explicitly, the $\lambda_4$ interaction term gives
\begin{align}
I_{H \Delta \bar\Delta} = 
v_4 \lambda_4 H \left[ \delta^{++} \delta^{--} + \delta^{+} \delta^{-} 
+ \frac{1}{2} ( \delta^{0^2} + \eta^{0^2}) \right] ~.
\label{eq:Hdd}
\end{align}

In this work, we consider the scenario where the DM particles annihilate through the $\chi$-$\bar\chi$-$H$ and $H$-$\Delta$-$\bar\Delta$ interactions given in Eqs.~(\ref{eq:igg}) and (\ref{eq:Hdd}), respectively, to a pair of Higgs triplet bosons.  To facilitate this pair annihilation process, we employ the Breit-Wigner enhancement through the relation $m_{H} \approx 2 m_\chi$, which affects both relic density~\cite{Gondolo:1990dk} and positoron/antiproton fluxes~\cite{Feldman:2008xs, Ibe:2008ye}.  Subsequently, we have $\delta^{\pm\pm} \to \ell^\pm{\ell'}^\pm$ and $\delta^\pm \to \ell^\pm \nu_{\ell'}$ dominantly and some small branching fractions of ${\cal O}(10^{-3})$ for $\delta^{\pm\pm} \to W^\pm W^\pm$, $\delta^\pm \to W^\pm Z$ and $\delta^0 \to ZZ/W^+W^-$, with details depending on the values of $v_\Delta$ and the Yukawa couplings ${\bf y}^{\ell}$.  As a result, the model can have simultaneous productions of $e^+$ and $W^+$, with some of the latter hadronizing into antiprotons, at significantly different rates.

Now the major free parameters in our analysis are $g_X$, $\lambda_4$, $m_\chi$, $m_H$, $M_\Delta$, $v_\Delta$, and ${\bf y}^{\ell}$.  In the DM annihilation amplitude, the gauge coupling $g_X$ and the quadruplet VEV $v_4$ always appear in the combination of product and can be rewritten in terms of $m_\chi$.  For the mass of $H$, we write
\begin{align}
m_{H} = 2 m_\chi (1 - \epsilon)
\label{eq:eps}
\end{align} 
and use $\epsilon$ as one positive free parameter of ${\cal O}(0.1)$ or smaller.  A Breit-Wigner enhancement is then obtained in the current DM annihilation rate because in the non-relativistic limit the annihilation cross section is written by
\begin{align}
 \sigma v \simeq
\frac{1}{192\pi} \left( \frac{\lambda_4}{m_\chi} \right)^2
\left[  
\left( \frac{v^2}{4} + 2 \epsilon \right)^2
+ \frac{\Gamma_H^2}{4m_\chi^2}
\right]^{-1}
\sqrt{1 - \frac{M_\Delta^2}{m_\chi^2}} ~,
\label{eq:sigmav}
\end{align}
where $\Gamma_H$ denotes the total width of $H$ and is found to be much smaller than $m_\chi$.  The average speed of DM $v$ in units of the speed of light is typically $\sim 10^{-3}$ at the current Universe and $\sim 0.3$ at the freeze-out.  Therefore, the dominant parameters in Eq.~(\ref{eq:sigmav}) are $\lambda_4$, $m_\chi$ and $\epsilon$.  The Yukawa couplings can be fixed by fitting to the neutrino mass measurements~\cite{PDG2014} and assuming the normal hierarchy as it is preferred by the positron flux excess~\cite{Chen:2014lla}.  Moreover, we have taken the CP-violating phase, the Majorana phases, and the mass of the lightest neutrino to be zero.  The conclusion will not change much if, for example, we take a tiny nonzero mass for the lightest neutrino.  Both lepton flavour-conserving and -violating channels have been included in the decays of the triplet Higgs bosons.  Our choice of $M_\Delta$ at the TeV scale exempts us from the constraint of the lower bound of about 400 GeV on the doubly-charged Higgs boson mass through the searches of like-sign dilepton channels~\cite{Chatrchyan:2012ya,ATLAS:2012hi}.  Otherwise, the results of our numerical analysis are not sensitive to $M_\Delta$ as long as it is smaller than $m_\chi$.  
In the following analysis, we take $M_\Delta = 800$ GeV as a reference value. 
In the end, we are left with four independent free parameters: $\lambda_4$, $m_\chi$, $\epsilon$, and $v_\Delta$.

\section{Numerical Analysis and Discussions}

We start by investigating the DM relic abundance $0.1172 \leq \Omega h^2 \leq 0.1204$ given by the PLANCK Collaboration at 90\% confidence level (CL)~\cite{Planck:2015xua}.  In our setup, the DM annihilation cross section 
$\sigma(\chi \bar\chi \to H \to \delta^{++}\delta^{--}, \delta^{+}\delta^{-}, \delta^0\delta^0, \eta\eta)$ is given by Eq.~(\ref{eq:sigmav}).  It is noted that the DM can also annihilate into a pair of SM particles ($W^+ W^-$, $ZZ$, $q \bar q$, etc) through the $h$-$H$ mixing, which is suppressed as we assume a small mixing angle $\theta$.  Our numerical analysis is done by utilizing the {\tt micrOMEGAs 4.1.5} package~\cite{Belanger:2014vza} implemented with the model to solve the Boltzmann equation for the DM relic density.

\begin{figure}[tbh]
\begin{center}
\includegraphics[width=90mm]{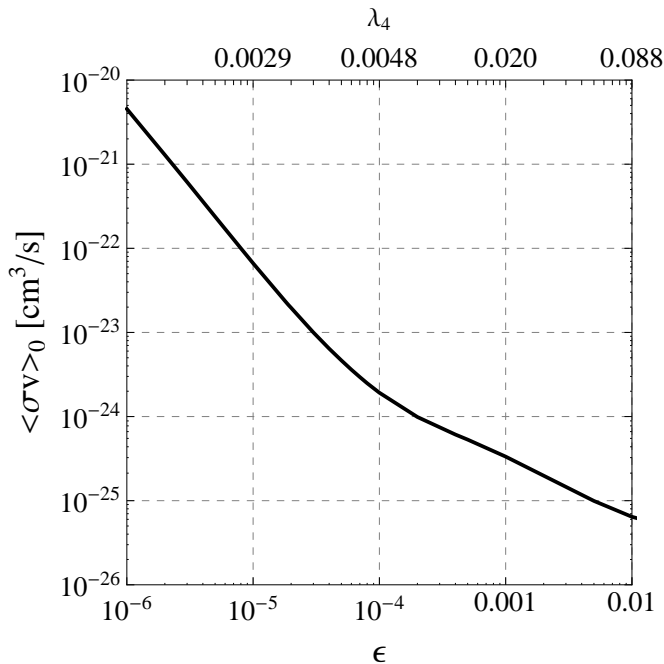}
\end{center}
\caption{Thermally averaged dark matter annihilation cross section at the current Universe as a function of $\epsilon$ in the lower horizontal axis, as required by the observed relic density.  The upper horizontal axis gives the corresponding value of $\lambda_4$.}
\label{fig:XS}
\end{figure}

Fig.~\ref{fig:XS} shows the thermally averaged cross section of DM annihilation to all triplet Higgs bosons, $\langle \sigma v \rangle_0$, at the present time for producing the right relic density.  It is shown as a function of $\epsilon$ defined in Eq.~(\ref{eq:eps}).  The corresponding values of $\lambda_4$ are indicated in the upper horizontal axis as well.  Although this curve does not have a sensitive dependence on the DM mass, we will fix $m_\chi = 2$ TeV because of better fits to the energy spectra of positron flux and antiproton flux ratio from the cosmic rays.  As seen in the plot, the cross section can reach $\sim 10^{-22}$ cm$^3$/s for $\epsilon \sim {\cal O}(10^{-5})$ because of the Breit-Wigner enhancement.  In this case, the coupling $\lambda_4 \simeq 3 \times 10^{-3}$.  As shown below, such a cross section is desirable for rendering the correct excess of the positron flux.

Again, using the {\tt micrOMEGAs 4.1.5} package~\cite{Belanger:2014vza}, we compute the positron and antiproton fluxes resulting from the decays of the charged Higgs bosons in the DM annihilation final states.  For the dark matter density profile, we take the NFW model with a local halo density of $0.3$ GeV/cm$^3$, a core radius of 20 kpc, and the distance from our solar system to the galactic center as $8.5$ kpc~\cite{Navarro:1996gj}.  For charged particle propagation through the space, we consider the three schemes MIN, MED, and MAX defined in Ref.~\cite{Donato:2003xg} to have the minimum, medium, and maximum charged particle flux, respectively.  The background for positron flux is provided by a fitting function given in Refs.~\cite{Baltz:1998xv,Baltz:2001ir}.  The background for antiproton flux is estimated by combining AMS proton flux data~\cite{Aguilar:2015ooa} and the flux ratio $\Phi_{\bar p}/\Phi_p$ estimated in Ref.~\cite{Giesen:2015ufa}.

\begin{figure}[tbh]
\begin{center}
\includegraphics[width=80mm]{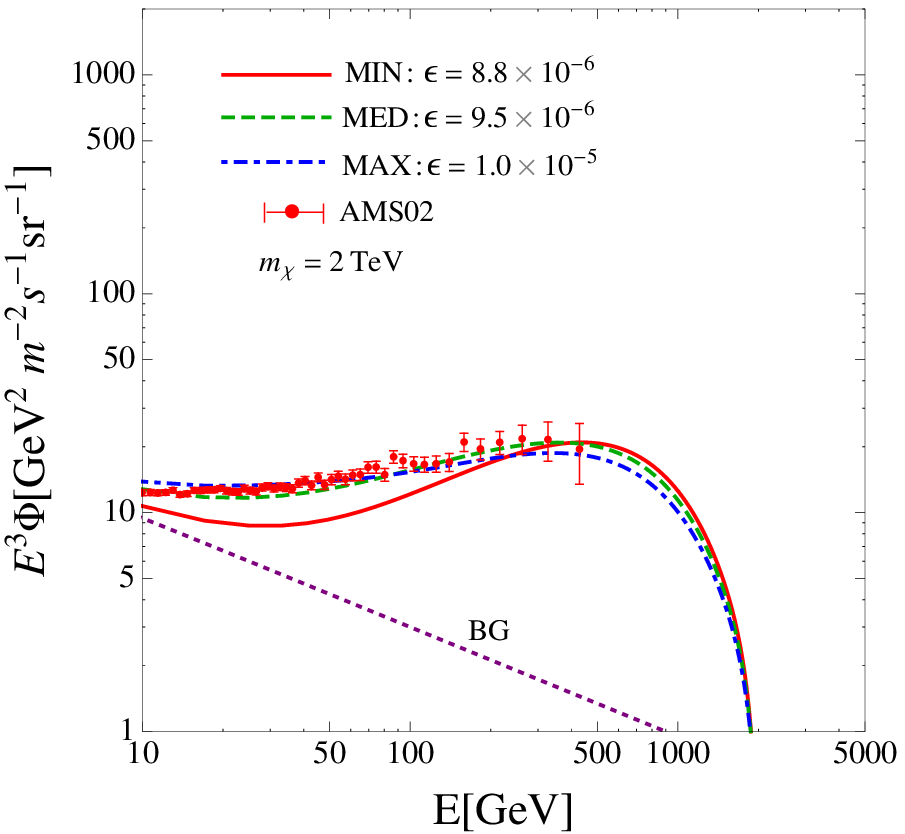}
\includegraphics[width=80mm]{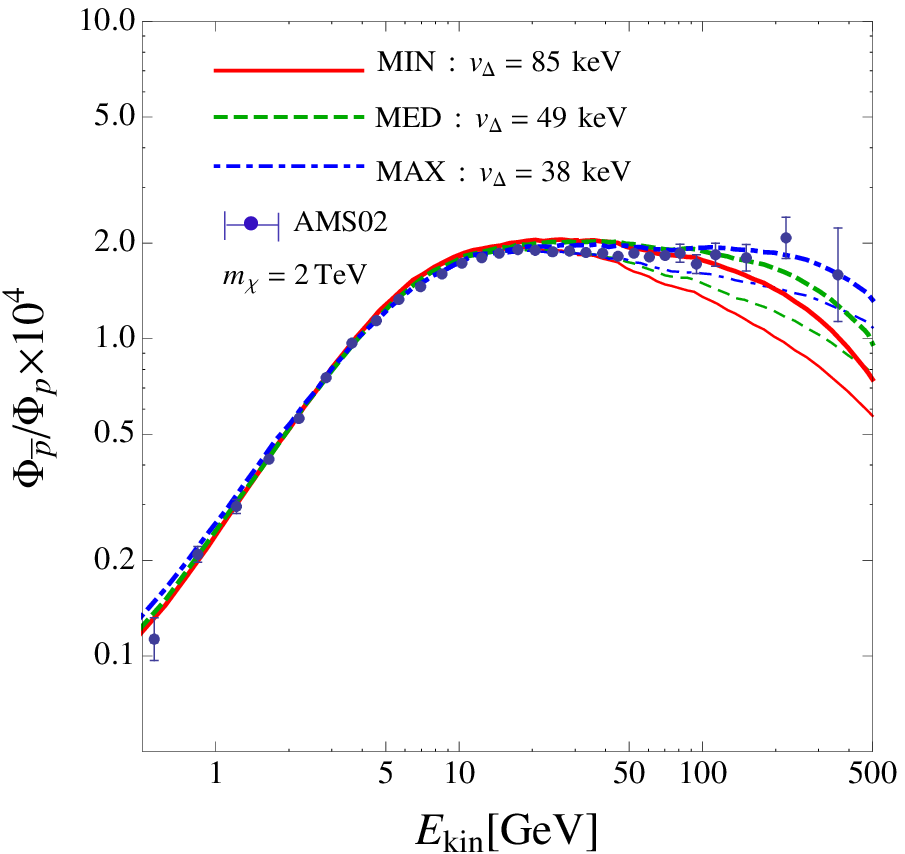}
\end{center}
\caption{Spectra of the positron flux (left) and the ratio of antiproton-to-proton fluxes (right) in comparison with those observed by AMS-02 from cosmic rays, drawn in thick curves.  Red solid curves are used for MIN, dashed green curves for MED, and dash-dotted blue curves for MAX.  Also indicated are the values of $\epsilon$ in the left plot and $v_\Delta$ in the right plot.  The background in the left plot is given by the purple dotted curve, and those in the right plot are given by the thin curves.}
\label{fig:fluxes}
\end{figure}

In Fig.~\ref{fig:fluxes}, the left plot shows the positron flux as a function of the positron energy.  The right plot shows the ratio of antiproton flux to proton flux as a function of the antiproton kinetic energy.  In both plots, we fix $m_\chi = 2$ TeV.  Qualitatively, changing the DM mass will shift the peaks of signals horizontally.  The magnitudes of signals are mainly controlled by $\epsilon$ for the positron flux and by $v_\Delta$ for the antiproton flux.  We take $\epsilon = 8.8 \times 10^{-6}$ and $v_\Delta = 85$ keV, corresponding to $Br(\delta^{\pm\pm} \to W^\pm W^\pm) = 4.3\%$, for the MIN scheme (red solid curves); $\epsilon = 9.5 \times 10^{-6}$ and $v_\Delta = 49$ keV, corresponding to $Br(\delta^{\pm\pm} \to W^\pm W^\pm) = 0.48\%$, for the MED scheme (green dashed curves); and $\epsilon = 1.0 \times 10^{-5}$ and $v_\Delta = 38$ keV, corresponding to $Br(\delta^{\pm\pm} \to W^\pm W^\pm) = 0.19\%$, for the MAX scheme (blue dash-dotted curves).

Here we make some comments regarding the constraint on DM annihilation cross section from the $\gamma$-ray flux provided from inclusive photon spectrum measurement~\cite{Ackermann:2012qk} and $\gamma$-ray observation of dwarf galaxies~\cite{Ackermann:2013yva} by Fermi-LAT.  Although DM annihilation to four-body final states, as in our case, are generally less constrained than two-body final states, our analysis assuming the normal hierarchy of neutrino masses has $\delta^{\pm \pm}$ decaying to energetic $\tau^\pm \tau^\pm$ and $\tau^\pm \mu^\pm$ at a branching fraction of around 30\% for both modes.  Such $\tau$ leptons in the final states may produce $\gamma$-rays subject to the Fermi-LAT constraint.  A detailed analysis with this taken into account will be presented in another work~\cite{CCN2015}.

Finally, we note in passing that the scattering cross section between the DM and the nucleons is small because the interactions between the DM and the SM particles is suppressed by the mixing angle $\theta$.  Therefore, the model can readily evade the constraints from current direct searches.

\section{Summary}

In the current work, we have constructed a model based on the Higgs triplet model and an additional $SU(2)_X$ gauge symmetry that links neutrino mass and dark matter physics.  With an $SU(2)_X$ quadruplet that develops a vacuum expectation value in its $T_3 = \pm 3/2$ component, the $SU(2)_X$ symmetry is broken down to a discrete $Z_3$ symmetry, thereby stabilising the $T_3 = \pm 1$ components of the gauge field as the dark matter candidates.  We have then analyzed the dark matter phenomenology of the model particularly in view of the recent AMS-02 data.  Through the physical scalar boson of the quadruplet as the messenger, the dark matter particles can annihilate primarily into a pair of triplet Higgs bosons, provided that the mixing between the quadruplet scalar particle and the SM-like Higgs boson is sufficiently small.  The correct dark matter relic abundance can be obtained when the mass of messenger is about twice that of the dark matter.  The charged triplet Higgs bosons in turn decay dominantly into leptonic final states to produce an excess of positrons.  However, there is a small branching fraction for the charged triplet Higgs bosons to decay into weak gauge bosons, part of which eventually hadronize into antiprotons.  The decay pattern of the charged triplet Higgs bosons is largely fixed by assuming the normal hierarchy and mass measurements of neutrinos.  We have studied three charged particle propagation schemes and found the corresponding parameters of the model that fit well with the positron flux and antiproton flux ratio observed by the AMS Collaboration.  Finally, we remark that the gauged $SU(2)_X$ for stable dark matter candidates in this study is only one choice.  Any DM model from a hidden sector that couples mainly to the triplet Higgs fields should have the same features as well.


\section*{Acknowledgments}

The authors would like to thank Yuan-Hann Chang for useful discussions about the AMS-02 data.  This work was supported in part by the Ministry of Science and Technology of Taiwan under Grant Nos.~MOST-103-2112-M-006-004-MY3 (CHC), MOST-100-2628-M-008-003-MY4 (CWC) and MOST-103-2811-M-006-030 (TN), as well as the World Premier International Research Center Initiative, Ministry of Education, Culture, Sports, Science and Technology, Japan.  We also thank the National Center for Theoretical Sciences for the support of useful facilities.


\end{document}